\documentclass{article}
\usepackage{authblk}
\usepackage{spconf, amsmath, graphicx}
\usepackage{multirow}
\usepackage{booktabs}
\usepackage{subfigure}
\usepackage{threeparttable}
\usepackage{amssymb}
\usepackage{cite}
\usepackage[T1]{fontenc}

\title{DUAL-PATH TRANSFORMER BASED NEURAL BEAMFORMER FOR TARGET SPEECH EXTRACTION}
\begin{document}
\ninept

\name{Aoqi Guo$^{\star \dagger}$\thanks{Work conducted when the first author was intern at SenseTime Research.} \qquad Sichong Qian$^{\dagger}$ \qquad Baoxiang Li$^{\dagger}$ \qquad Dazhi Gao$^{\star}$\thanks{Dazhi Gao is the corresponding author.}}
\address{$^{\star}$ Ocean University of China, Qingdao, China\\
	$^{\dagger}$ SenseTime Research, Beijing, China}

\maketitle
\begin{abstract}
Neural beamformers, which integrate both pre-separation and beamforming modules, have demonstrated impressive effectiveness in target speech extraction. Nevertheless, the performance of these beamformers is inherently limited by the predictive accuracy of the pre-separation module. In this paper, we introduce a neural beamformer supported by a dual-path transformer. Initially, we employ the cross-attention mechanism in the time domain to extract crucial spatial information related to beamforming from the noisy covariance matrix. Subsequently, in the frequency domain, the self-attention mechanism is employed to enhance the model's ability to process frequency-specific details. By design, our model circumvents the influence of pre-separation modules, delivering performance in a more comprehensive end-to-end manner. Experimental results reveal that our model not only outperforms contemporary leading neural beamforming algorithms in separation performance but also achieves this with a significant reduction in parameter count.
\end{abstract}
\section{Introduction}
\label{sec:intro}

Neural beamformers have demonstrated exceptional capabilities in the field of multi-channel target speech extraction \cite{zmolikova2023neural}. One of the pioneering approaches that combines deep learning with traditional beamforming algorithms is the mask-based neural beamformer\cite{7471664}. This architecture typically consists of a pre-separation module, followed by a beamforming module. Initially, the pre-separation module generates a series of time-frequency masks\cite{erdogan2016improved,higuchi2016robust,williamson2015complex,mack2019deep}. These masks are then used to calculate the covariance matrices for either the speech or noise signals, as required by the beamforming algorithm. Subsequently, these covariance matrices are fed into classical beamforming algorithms to calculate the beamforming weights. Given that traditional beamforming algorithms are typically deterministic and operate independently of the pre-separation module, any fluctuation in the output accuracy of the pre-separation module can have a substantial impact on the system's performance. To address this, newer models such as ADL-MVDR \cite{zhang2021adl}, GRNNBF \cite{xu2021generalized} and SARNN \cite{li2021mimo} started integrating neural networks into the beamforming module. This enhances the synergy between the pre-separation and beamforming modules, thereby improving the overall performance of speech separation algorithms. However, the predictive accuracy of the pre-separation module remains a limiting factor in the overall effectiveness of the system.

Recently, there have been further developments in the integration of neural networks and beamforming algorithms. Researchers have developed models that utilize neural networks to directly characterize signals and predict beamforming weights in either the time or frequency domain. For instance, EABNet \cite{li2022embedding} directly models the time-frequency characteristics of signals captured by microphone arrays, aiming to capture more comprehensive information than mere covariance matrices in order to predict beamforming weights directly. On a parallel note, Yi Luo et al. introduced FasNet-TAC \cite{luo2020end}, which performs filtering and summation operations in the time domain. By utilizing a neural network to model input features, the model's coherence is improved. Nevertheless, it is important to note that the fundamental concept of beamforming revolves around spatial domain signal filtration. The way in which these models handle spatial signal information remains less transparent. As a result, they exhibit reduced interpretability and robustness when compared to neural beamformers that are based on spatial information obtained from covariance matrices.

In this paper, we introduce a neural beamformer that utilizes a dual-path transformer. Initially, we model both the spatial features of the input and the noisy covariance matrix. We then implement a cross-attention mechanism \cite{vaswani2017attention} at the narrowband level to extract spatial information that is relevant to beamforming from the noisy covariance matrix, utilizing spatial features as cues. Following this, we employ a self-attention mechanism at the broadband level, enhancing the model's capability to capture inter-frequency relationships. Ultimately, we derive beamforming weights directly from the spatial information that we have modeled. Our approach avoids the estimation of intermediate variables, and facilitates a more end-to-end prediction of beamforming weights. Most notably, our model maintains a level of interpretability while significantly reducing the total number of parameters.

The remainder of this paper is organized as follows. Section 2 presents the signal model and the theoretical basis for our proposed model. Section 3 details our proposed neural beamforming algorithm. Section 4 provides an overview of the experimental setup and presents an analysis of the experimental results. Finally, Section 5 concludes the paper.

\section{Signal Models and Methods}
\label{sec:format}

Consider the far-field frequency-domain signal model in the real scene, described as
\begin{equation}
	Y(t,f) = X(t,f) + S(t,f) + N(t,f)
\end{equation}
where $Y(t,f)=[Y^{(0)}(t,f), Y^{(1)}(t,f),..., Y^{(M-1)}(t,f)]^{T}$ indicates the frequency-domain signal received by the M-channel microphone array. $X(t,f)$, $S(t,f)$ are the reverberated speech signals of the target speaker and the interference speaker respectively, and $N(t,f)$ represents the background noise. When not focusing on the dereverberation task, the task goal of multi-channel target speech extraction is to extract the monaural speech $X^{'}(t,f)$ of the target speaker from the noisy signal $Y(t,f)$.

The purpose of the beamforming algorithm is to obtain the filter weight $w$ for the array observation signal, and extract the desired signal by employing spatial filtering, that is:
\begin{equation}
	X^{'}(t,f) = w^H\cdot Y(t,f)
\end{equation}
where $(\cdot)^H$ is the hermitian transpose, and \textquotedbl$\cdot$\textquotedbl denotes the matrix multiplication. Typically, beamforming algorithms calculate the covariance matrix for speech or noise signals to determine beamforming weights. Hence, the prediction accuracy of the covariance matrix has a great influence on the overall performance.

Generally, considering the covariance matrix of the noisy signal:
\begin{equation}
	\Phi_{YY}(t,f)=Y(t,f)\cdot Y^{H}(t,f) \in \mathbb{C}^{M\times M}
\end{equation}
when the speech and interference noise signals are independent of each other and the influence of background noise is ignored, we can get:
\begin{equation}
	\Phi_{YY}(t,f) \approx \Phi_{XX}(t,f) + \Phi_{NN}(t,f)
\end{equation}
It is evident that the covariance matrix of the noisy signal encompasses all the information from the covariance matrices of both speech and noise signals. With this understanding, we employ a neural network to model the noisy covariance matrix and input features, thereby predicting the beamforming weights, described as:
\begin{equation}
	w(t,f) = Model(Features(t,f), \Phi_{YY}(t,f))
\end{equation}

\section{Our Proposed Neural Beamformer}
\label{sec:pagestyle}
Fig. 1 illustrates the overall structure of our proposed model. Initially, the model computes the input spatial features as well as the noisy covariance matrix. These are then transformed to a uniform dimensionality via a DNN-RNN architecture. Finally, the beamforming weights are predicted using a dual-path transformer. The specific steps are as follows:

\subsection{Feature Extraction}
We select the first channel of the microphone array as the reference channel and compute the magnitude spectrum to serve as the input feature.
\begin{equation}
	Y^{(0)}_{mag}(t,f) = |Y^{(0)}(t,f)|
\end{equation}
We then choose P pairs of microphones to calculate the phase difference between each pair, which serves as a spatial feature.
\begin{equation}
	\begin{split}
		cosIPD_{p}(t,f) = cos(\angle Y_p^{1}(t,f)-\angle Y_p^{0}(t,f))
	\end{split}
\end{equation}
while $\angle(\cdot)$ outputs the angle of the input argument. $Y_p^{0}(t,f)$ and $Y_p^{1}(t,f)$ represent the spectrum of the signal received by each microphone in the p-th microphone pair, respectively.
Ultimately, we compute the angle feature \cite{8639593}, given that the Direction of Arrival(DOA) of the target speaker is known.
\begin{equation}
	AF_{\theta}(t,f) = \sum_{p=1}^P cos(IPD_{p}(t,f)-\Delta_{\theta,p}(t,f))
\end{equation}
where $\Delta_{\theta,p}(t,f)$ represents the ground truth phase difference given the direction of arrival $\theta$ and the p-th microphone pair, and for a speaker at a fixed position, its characteristics remain consistent across all time frames.
All these features are stacked along the channel dimension to serve as the model's input features, that is $Features(t,f) \in \mathbb{R}^{P+1+1}$. After computing the complex-valued covariance matrix of the noisy signal through Eq. (3), we concatenate the real and imaginary components along the channel dimension, that is
\begin{equation}
	\Phi(t,f) =[\Phi^{r}_{YY}(t,f), \Phi^{i}_{YY}(t,f)] \in \mathbb{R}^{M\times M\times2}
\end{equation}
where $\Phi^{r}_{YY}(t,f)$ and $\Phi^{i}_{YY}(t,f)$ represent the real and imaginary components, respectively.

\subsection{Dimensional modeling}
We first process each frequency point independently at the narrowband level and then transform the input features and noisy covariance matrix to the same dimension through two different Conv1D layers:
\begin{equation}
	E_{Feat}(t,f) = Conv1D_{1}(Features(t,f)) \in \mathbb{R}^{D}
\end{equation}
\begin{equation}
	E_{\Phi}(t,f) = Conv1D_{2}(\Phi(t,f)) \in \mathbb{R}^{D}
\end{equation}
where $D$ represents the embedding dimensions.
Then they are concatenated along the channel dimension, and then input to the GRU to enhance the modeling and mapping capabilities of the network.
\begin{equation}
	E_{mix}(t,f) = [E_{Feat}(t,f), E_{\Phi}(t,f)] \in \mathbb{R}^{D\times2}
\end{equation}
\begin{equation}
	E^{'}_{mix}(t,f) = GRU(E_{mix}(t,f)) \in \mathbb{R}^{D\times2}
\end{equation}
After splitting the modeled $E^{'}_{mix}(t,f)$, we get $E^{'}_{Feat}(t,f)\in \mathbb{R}^{D}$ and $E^{'}_{\Phi}(t,f)\in \mathbb{R}^{D}$, which represents the spatial information of the input features and the noisy covariance matrix, respectively.
\begin{equation}
	[E^{'}_{Feat}(t,f), E^{'}_{\Phi}(t,f)] = Chunk(E^{'}_{mix}(t,f))
\end{equation}

\begin{figure*}[t]
	\begin{center}
		\includegraphics[width=180mm]{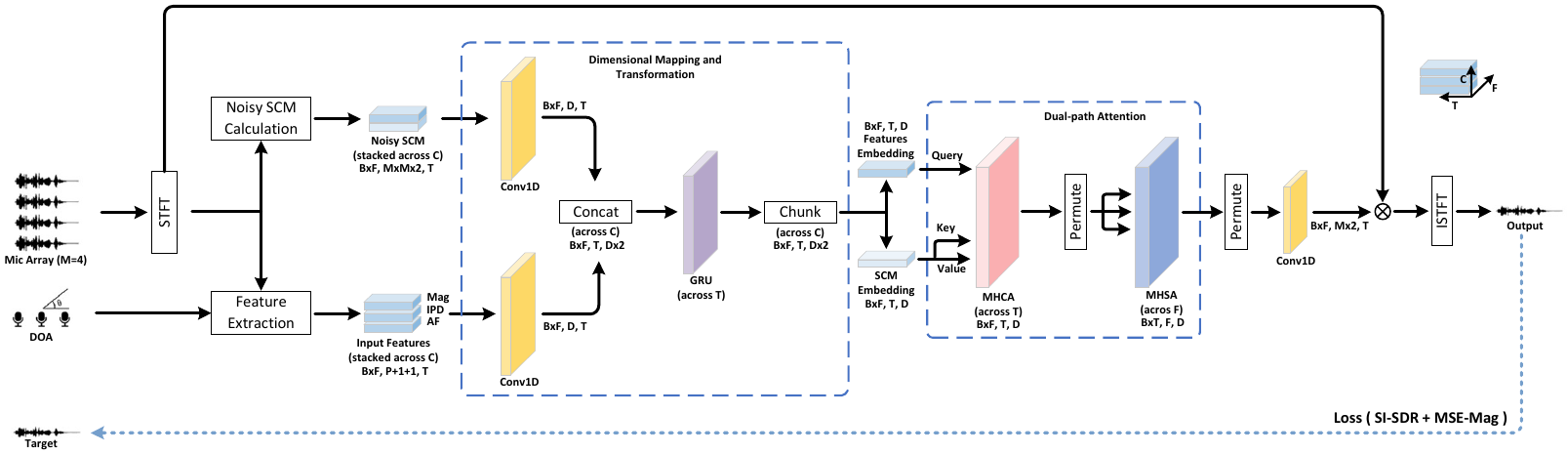}
	\end{center}
	\caption{The overall structure of our proposed model. MHCA is a time-domain multi-head cross-attention module, and MHSA is a frequency-domain multi-head self-attention module. The MHCA and Conv1D layers process each frequency independently at the narrowband level. The MHSA layer models frequency-domain information frame by frame at wideband level. B, C, F, T, D and M represent batch size, channel dimensions, frequency dimensions, time dimensions, embedding dimensions and channels, respectively.}
\end{figure*}

\begin{figure}[htbp]
	\begin{center}
		\includegraphics[width=70mm]{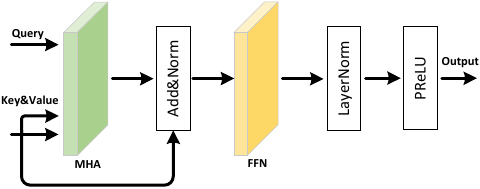}
	\end{center}
	\caption{Structural diagram of the attention module. At the narrowband level, the spatial information is used as Query, and the covariance matrix is used as Key and Value. At the broadband level, Query, Key, and Value are all embedding itself.}
\end{figure}

\subsection{Dual-path Transformer Based Module}
Initially, we leverage a cross-attention module in the time domain to extract beamforming-related information. We use the modeled input features, $E^{'}_{Feat}(t,f)$, as the query, and the noisy covariance matrix, $E^{'}_{\Phi}(t,f)$, as both key and value for performing the operation of the cross-attention mechanism. This approach enables us to focus on beamforming-relevant spatial information.

\begin{equation}
	E_{w}(t,f) = MHCA(E^{'}_{Feat}(t,f), E^{'}_{\Phi}(t,f), E^{'}_{\Phi}(t,f))
\end{equation}

The prior operations of the model are executed independently at each frequency. To leverage inter-frequency information more effectively, we transform the dimension of the embedding and employ a self-attention module in the frequency domain, and apply it frame by frame at the broadband level.
\begin{equation}
	E^{'}_{w}(t,f) = MHSA(E_{w}(t,f), E_{w}(t,f), E_{w}(t,f))
\end{equation}
Following these steps, the model has extracted beamforming-related information from the noisy covariance matrix. Ultimately, we employ a Conv1D layer to predict the real and imaginary parts of the beamforming weights, thereby enabling the application of beamforming to the array observation signals.
\begin{equation}
	w(t,f)=Conv1D_{3}(E^{'}_{w}(t,f))
\end{equation}
The time-domain speech signal is restored after the processed frequency spectrum is subjected to Inverse Short-Time Fourier Transform(ISTFT). Note that only the target speech extraction is considered and dereverberation is not addressed in this paper.

\section{Dataset and Experimental Setup}
\label{sec:typestyle}
\subsection{Dataset}
The source speech signals for target and interfering speakers are obtained from AISHELL-1 \cite{bu2017aishell}, with background noise data sourced from DNS2023 \cite{dubey2023icassp}. For indoor reverberation scenes, we randomly set room dimensions between [3, 3, 1.5] and [8, 8, 2.5] meters, with a reverberation time (rt60) of 0.1 to 0.6 seconds. The simulation uses the image source method, producing room impulse responses for target and interference signals\cite{scheibler2018pyroomacoustics}. The microphone array consists of a four-element linear array with 3 cm spacing. To maintain spatial independence, we set a minimum 5° angle between the two sources relative to the microphone array. We adjust the Signal-Interference Ratio (SIR) between target and interference signals from [-6, 6] dB, and add background noise from [-5, 20] dB to enhance model robustness. During training, all data is divided into 4 seconds. This process yields approximately 133.3 hours of training data (120,000 pieces), 15.6 hours of validation data (14,000 pieces), and 7.8 hours of test data (7,000 pieces), all down-sampled to 16kHz.
\begin{table*}[htbp]
	\centering
	\begin{threeparttable}
		\caption{PESQ, STOI, Si-SDR and WER of several baselines and the proposed DPTBF model.}
		\begin{tabular}{p{10.125em}cccccc}
			\toprule[1pt]
			\multirow{1.5}[2]{*}{Systems} & \multicolumn{1}{p{3.875em}}{GMACs} & \multicolumn{1}{c}{\multirow{1.5}[2]{*}{Para.(M)}} & \multicolumn{4}{c}{Simulated Data} \\
			\multicolumn{1}{c}{} & \multicolumn{1}{p{3.875em}}{(per sec.)} &       & \multicolumn{1}{p{2.8em}}{PESQ\bf↑} & \multicolumn{1}{p{2.5em}}{STOI\bf↑} & \multicolumn{1}{p{3.7em}}{Si-SDR\bf↑} & \multicolumn{1}{p{3.8em}}{WER(\%)\bf↓} \\
			\midrule
			Reverberant Clean & $-$     & $-$     & 4.5   & 1.0     & $\infty$  & 2.25 \\
			No processing & $-$     & $-$     & 1.148 & 0.563 & -1.76 & 68.02 \\
			\midrule
			IRM MVDR \cite{7471664} & 0.35  & 5.33   & 1.586 & 0.757 & 5.25  & 25.13 \\
			MISO Conv-TasNet \cite{gu2020multi} & \bf0.33  & 5.4   & 1.566 & 0.759 & 5.96  & 28.15 \\
			GRNNBF \cite{xu2021generalized} & 50.17 & 15.73 & 2.176 & 0.845 & 8.42  & 13.15 \\
			DPTBF (proposed) & 13.52 & 0.96  & \bf2.313 & \bf0.861 & \bf9.34  & \bf9.45 \\
			\midrule
			DPTBF (less) & 3.44  & \bf0.24  & 2.244 & 0.855 & 8.96  & 12.52 \\
			-FA   & 13.24 & 0.88  & 2.096 & 0.83  & 8.25  & 14.33 \\
			+SC   & 13.52 & 0.96  & 2.242 & 0.854 & 8.96  & 12.44 \\
			\bottomrule[1pt]
		\end{tabular}%
		\begin{tablenotes}[para,flushleft]
			\item "FA" means Frequency-domain Attention module and "SC" means Skip Connection for GRU.
		\end{tablenotes}
		\label{tab:addlabel}%
	\end{threeparttable}
\end{table*}%

\subsection{Implementation Details}
During the training process, a 512-point STFT is utilized to extract audio features using a 32ms Hann window with a 50\% overlap. We select three microphone pairs(P=3), specifically (0,1), (0,2) and (0,3), to calculate spatial features. The unidirectional GRU layer includes 256 hidden layer units, while the cross-attention and self-attention module dimensions are set at 128(D=128). The final linear layer predicts complex-valued beamforming weights, hence the output dimension is configured to 8(M$\times$ 2). More details of the model are provided here.\footnote{https://github.com/Aworselife/DPTBF}

The network undergoes 60 epochs of training with a batch size of 20. We use the Adam optimizer with an initial learning rate of 2e-3, decaying exponentially at 0.98 per epoch. A maximum gradient clipping of 10 accelerates network convergence.

We use the MIMO Conv-TasNet \cite{gu2020multi}, IRM MVDR\cite{7471664} and GRNNBF \cite{xu2021generalized} models as our experimental baselines for comparison. In line with the settings from the original paper, we employ a 3x8 Temporal Convolutional Network(TCN)\cite{bai2018empirical} Block for the MIMO Conv-TasNet to model input features, which predicts the Complex-Ratio-Mask(CRM)\cite{williamson2015complex} to restore the spectrum of the reference channel. Concurrently, using the same configuration as MIMO Conv-TasNet, we predict the Ideal-Ratio-Mask (IRM)\cite{higuchi2016robust} for both speech and noise signals and integrate it into the mask-based MVDR for computation. For the GRNNBF, a 4x8 TCN block is used to estimate the Complex-Ratio-Filter (CRF)\cite{schroter2022deepfilternet}. This prediction aids in calculating the covariance matrices for both speech and noise. This matrix is subsequently fed into a GRU layer comprising 500 hidden units to predict the beamforming weights. For our loss function, we combine the Si-SDR \cite{8683855} of the time-domain signal with the MSE of the magnitude spectrum, assigning them equal weights to derive the composite loss function.

\subsection{Experiment Results}
Table 1 presents a comparison between our proposed DPTBF model, the baseline model, and the results from ablation experiments. We evaluate the performance of the model using PESQ\cite{941023}, STOI\cite{5495701}, Si-SDR, and measure the Word Error Rate (WER) using the ASR model from a specific source\cite{gao2022paraformer}. The IRM MVDR model experiences a significant performance drop under conditions of background noise and strong reverberation due to the inherent residual noise issue in MVDR. The MISO Conv-TasNet, which predicts the target speech mask using TCN, eliminates beamforming operations and reduces computational load. However, it does come with a large number of parameters due to the use of multi-layer stacked TCN blocks. Building on the stacked TCN block, GRNNBF integrates an RNN for covariance matrix modelling, leading to a considerable increase in the parameter count. In terms of WER, MISO Conv-TasNet delivers the worst performance, primarily due to the significant spectral distortion it induces. Although the no-distortion constraint of IRM MVDR ensures some level of performance, it does not fully achieve the desired effect due to notable residual noise. In contrast, our proposed DPTBF significantly reduces parameters and computations compared to the baseline models, while also enhancing performance.

We further refined our model into a more streamlined version of DPTBF by reducing GRU hidden layer units from 256 to 128. Even though there is a performance decrease with fewer parameters, the experimental results remain encouraging. Ablation experiments on DPTBF show that using a frequency-domain self-attention module enhances the model's capability to capture frequency-domain information, thereby improving the performance of separation. However, adding skip connections to the GRU layer, which models spatial information and noisy covariance matrix information, did not improve the performance. This could potentially be due to the network's shallow depth and minimal loss of information.

\subsection{Spectrogram Analysis}
We selected a challenging audio sample from the test set to evaluate the performance of different models in challenging conditions. This sample had a low signal-to-noise ratio and a small angle between the two speakers. The noise signal received by the reference channel is depicted in Fig. 3(a), while the reverberated speech of the target speaker is illustrated in Fig. 3(f). Both the baseline and DPTBF models processed this data, with their results presented in Fig. 3(b-e). The IRM MVDR model, limited by its inherent algorithm, produces speech output with substantial residual noise. The MISO Conv-TasNet model, which utilizes TCN for direct mask prediction, shows robust noise suppression but suffers from significant spectral distortion, which will seriously affect the recognition results of the ASR model. On the other hand, the GRNNBF model has less spectral distortion, but it still struggles with residual noise. In comparison to these baseline models, our proposed DPTBF model excels in reducing interfering noise, mitigating background noise, and improving spectral distortion.
\begin{figure}[t]
	\subfigure[Noisy, Si-SDR: -3.4dB]{
		\centering
		\includegraphics[width=0.23\textwidth]{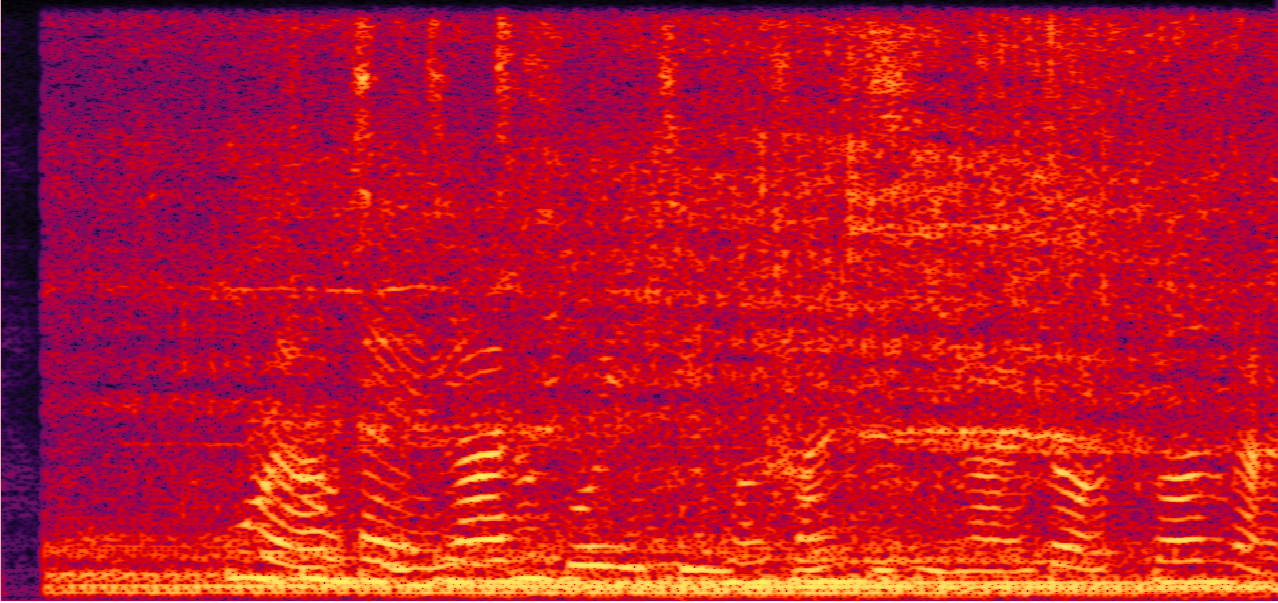}
	}\subfigure[IRM MVDR, Si-SDR: 2.02dB]{
		\centering
		\includegraphics[width=0.23\textwidth]{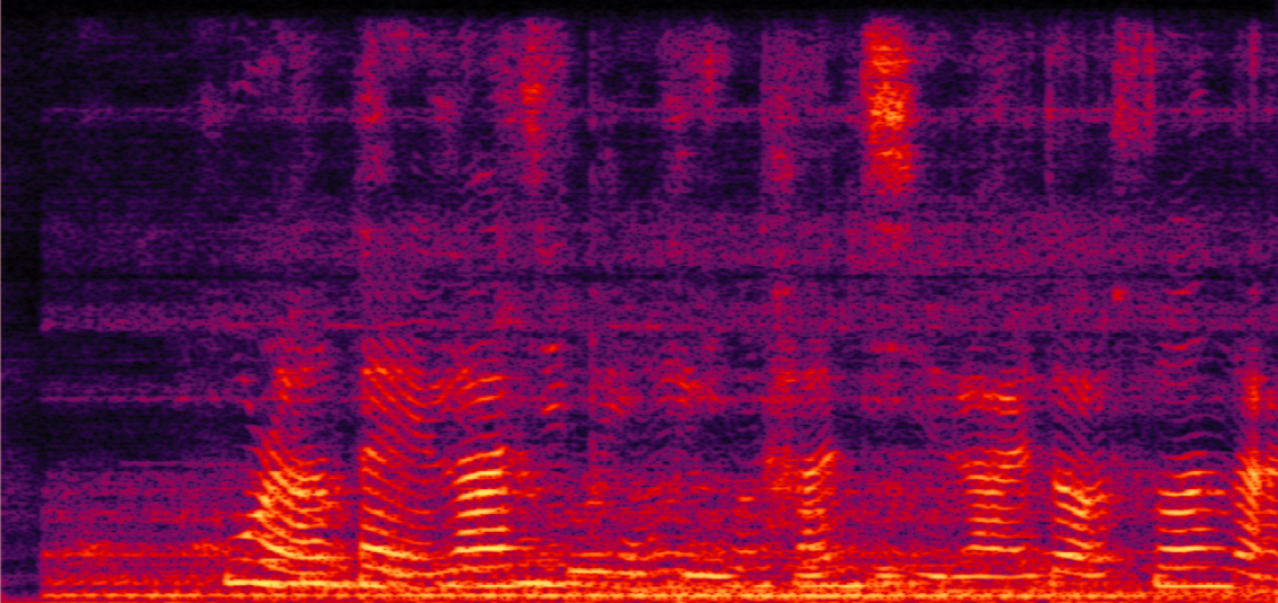}
	}
	
	\subfigure[Conv-TasNet, Si-SDR: 2.44dB]{
		\centering
		\includegraphics[width=0.23\textwidth]{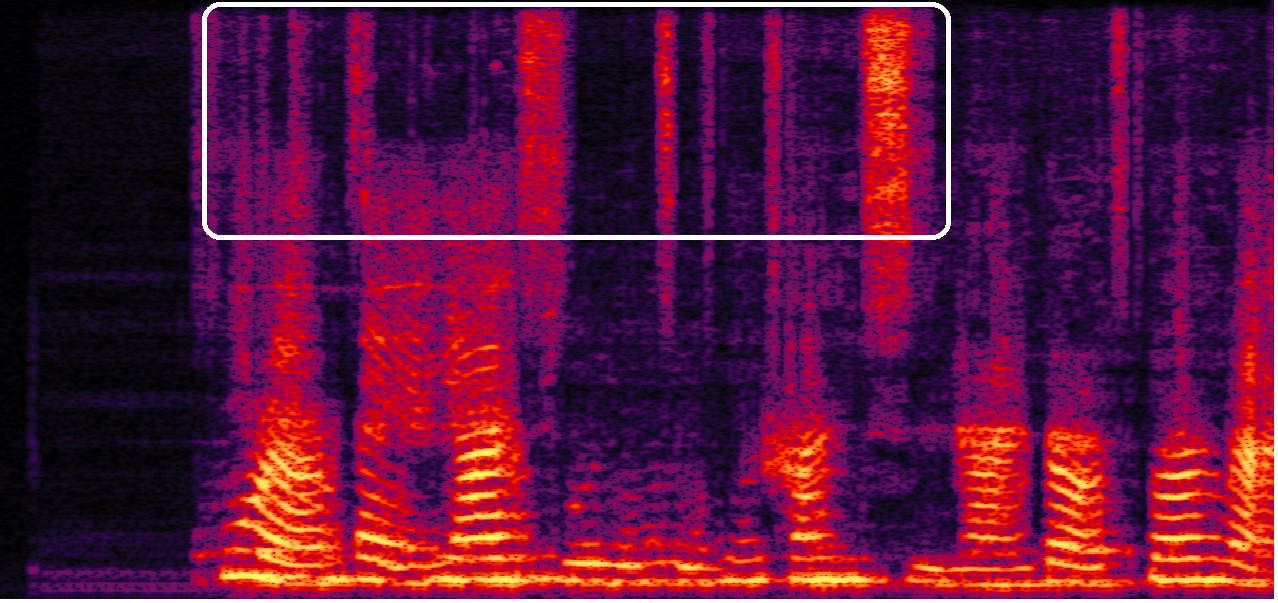}
	}\subfigure[GRNNBF, Si-SDR: 6.91dB]{
		\centering
		\includegraphics[width=0.23\textwidth]{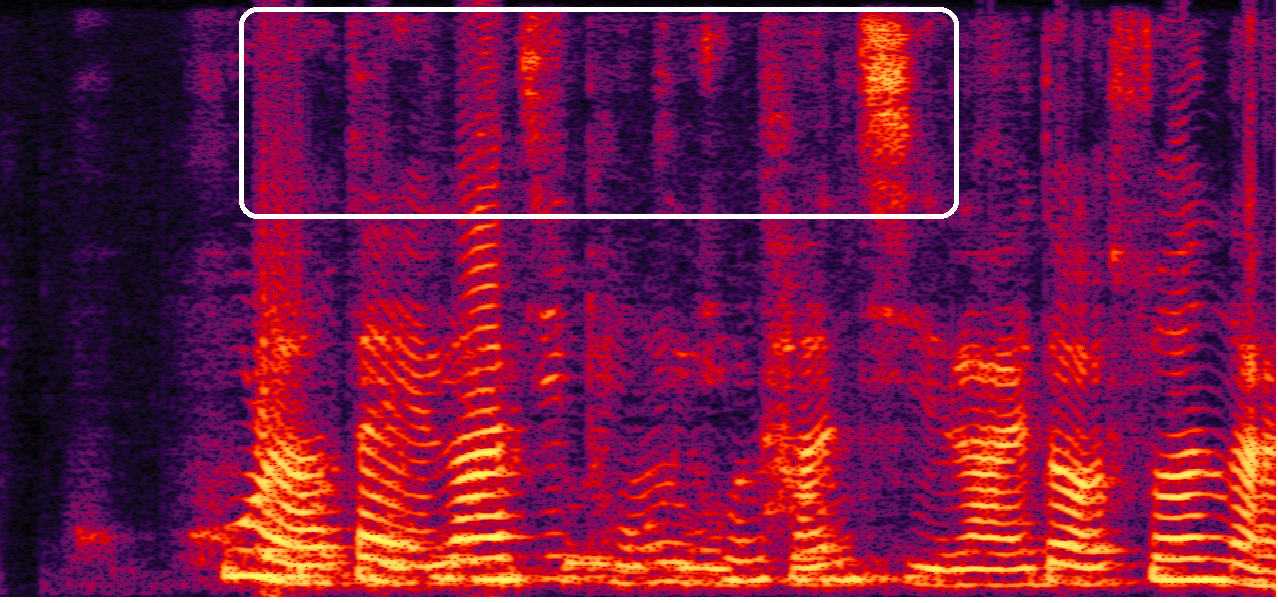}
	}
	
	\subfigure[Proposed, Si-SDR: 8.67dB]{
		\centering
		\includegraphics[width=0.23\textwidth]{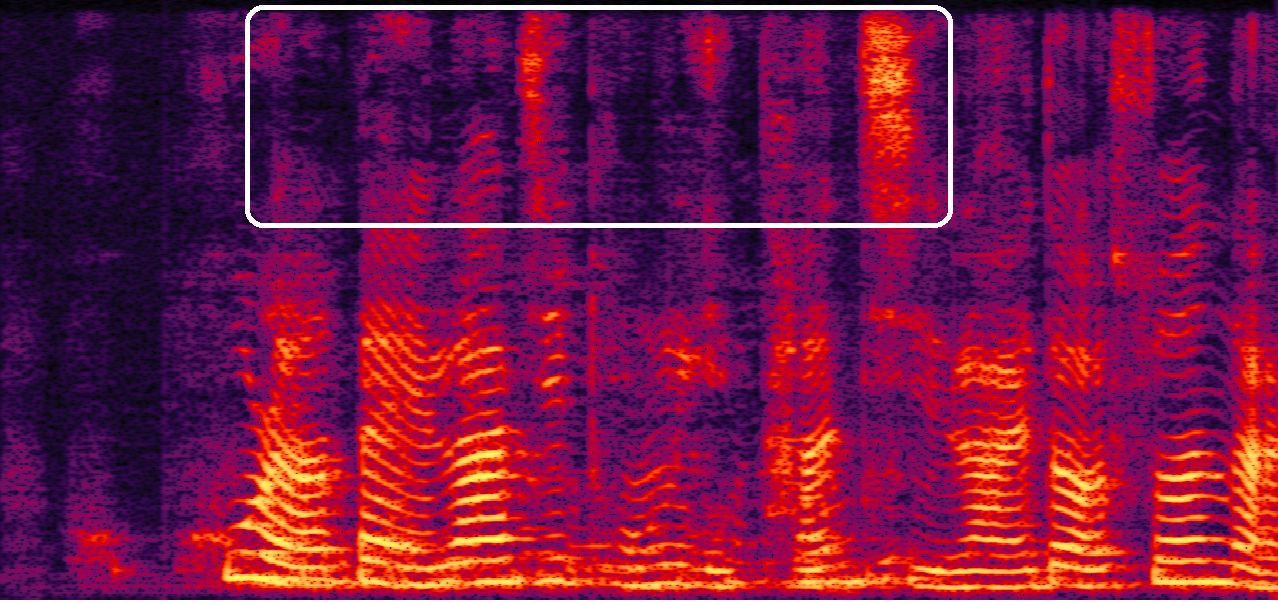}
	}\subfigure[Target]{
		\centering
		\includegraphics[width=0.23\textwidth]{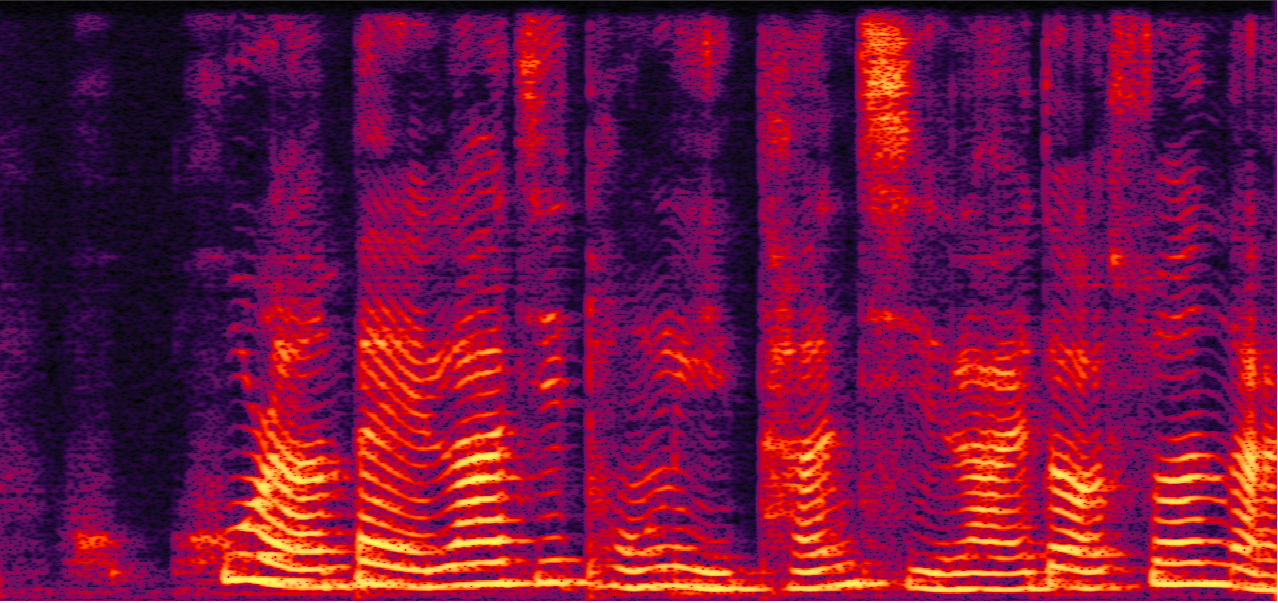}
	}\caption{Spectrum of a piece of data in the test set after processing. SIR=-2.2dB, SNR=-4.8dB. The angle between the speaker is 9°.\label{inut_output}}
\end{figure}

\section{Conclusions}
\label{ssec:subhead}

In summary, we propose a neural beamformer based on a dual-path transformer architecture. By incorporating both cross-attention and self-attention mechanisms, our model efficiently eliminates the need to estimate intermediate variables and overcomes the limitations of pre-separation modules. Empirical results underscore the superior performance of the model in target speech extraction tasks, further confirming the effectiveness of the attention mechanism in extracting spatial information relevant to beamforming from the covariance matrix. Looking ahead, our goal is to reduce the computational complexity of the model without compromising separation effectiveness, while also expanding the model to support the multiple-input multiple-output paradigm for additional speakers.


\vfill\pagebreak
\bibliographystyle{IEEEbib}
\bibliography{mybib}

\begin{thebibliography}{10}

\bibitem{zmolikova2023neural}
Katerina Zmolikova, Marc Delcroix, Tsubasa Ochiai, Keisuke Kinoshita, Jan
  {\v{C}}ernock{\`y}, and Dong Yu,
\newblock ``Neural target speech extraction: An overview,''
\newblock {\em IEEE Signal Processing Magazine}, vol. 40, no. 3, pp. 8--29,
  2023.

\bibitem{7471664}
Jahn Heymann, Lukas Drude, and Reinhold Haeb-Umbach,
\newblock ``Neural network based spectral mask estimation for acoustic
  beamforming,''
\newblock in {\em 2016 IEEE International Conference on Acoustics, Speech and
  Signal Processing (ICASSP)}, 2016, pp. 196--200.

\bibitem{erdogan2016improved}
Hakan Erdogan, John~R Hershey, Shinji Watanabe, Michael~I Mandel, and Jonathan
  Le~Roux,
\newblock ``Improved mvdr beamforming using single-channel mask prediction
  networks.,''
\newblock in {\em Interspeech}, 2016, pp. 1981--1985.

\bibitem{higuchi2016robust}
Takuya Higuchi, Nobutaka Ito, Takuya Yoshioka, and Tomohiro Nakatani,
\newblock ``Robust mvdr beamforming using time-frequency masks for
  online/offline asr in noise,''
\newblock in {\em 2016 IEEE International Conference on Acoustics, Speech and
  Signal Processing (ICASSP)}. IEEE, 2016, pp. 5210--5214.

\bibitem{williamson2015complex}
Donald~S Williamson, Yuxuan Wang, and DeLiang Wang,
\newblock ``Complex ratio masking for monaural speech separation,''
\newblock {\em IEEE/ACM transactions on audio, speech, and language
  processing}, vol. 24, no. 3, pp. 483--492, 2015.

\bibitem{mack2019deep}
Wolfgang Mack and Emanu{\"e}l~AP Habets,
\newblock ``Deep filtering: Signal extraction and reconstruction using complex
  time-frequency filters,''
\newblock {\em IEEE Signal Processing Letters}, vol. 27, pp. 61--65, 2019.

\bibitem{zhang2021adl}
Zhuohuang Zhang, Yong Xu, Meng Yu, Shi-Xiong Zhang, Lianwu Chen, and Dong Yu,
\newblock ``Adl-mvdr: All deep learning mvdr beamformer for target speech
  separation,''
\newblock in {\em ICASSP 2021-2021 IEEE International Conference on Acoustics,
  Speech and Signal Processing (ICASSP)}. IEEE, 2021, pp. 6089--6093.

\bibitem{xu2021generalized}
Yong Xu, Zhuohuang Zhang, Meng Yu, Shi-Xiong Zhang, and Dong Yu,
\newblock ``Generalized spatial-temporal rnn beamformer for target speech
  separation.,''
\newblock in {\em Interspeech}, 2021, pp. 3076--3080.

\bibitem{li2021mimo}
Xiyun Li, Yong Xu, Meng Yu, Shi-Xiong Zhang, Jiaming Xu, Bo~Xu, and Dong Yu,
\newblock ``Mimo self-attentive rnn beamformer for multi-speaker speech
  separation,''
\newblock in {\em Interspeech}, 2021, pp. 1119--1123.

\bibitem{li2022embedding}
Andong Li, Wenzhe Liu, Chengshi Zheng, and Xiaodong Li,
\newblock ``Embedding and beamforming: All-neural causal beamformer for
  multichannel speech enhancement,''
\newblock in {\em ICASSP 2022-2022 IEEE International Conference on Acoustics,
  Speech and Signal Processing (ICASSP)}. IEEE, 2022, pp. 6487--6491.

\bibitem{luo2020end}
Yi~Luo, Zhuo Chen, Nima Mesgarani, and Takuya Yoshioka,
\newblock ``End-to-end microphone permutation and number invariant
  multi-channel speech separation,''
\newblock in {\em ICASSP 2020-2020 IEEE International Conference on Acoustics,
  Speech and Signal Processing (ICASSP)}. IEEE, 2020, pp. 6394--6398.

\bibitem{vaswani2017attention}
Ashish Vaswani, Noam Shazeer, Niki Parmar, Jakob Uszkoreit, Llion Jones,
  Aidan~N Gomez, {\L}ukasz Kaiser, and Illia Polosukhin,
\newblock ``Attention is all you need,''
\newblock {\em Advances in neural information processing systems}, vol. 30,
  2017.

\bibitem{8639593}
Zhuo Chen, Xiong Xiao, Takuya Yoshioka, Hakan Erdogan, Jinyu Li, and Yifan
  Gong,
\newblock ``Multi-channel overlapped speech recognition with location guided
  speech extraction network,''
\newblock in {\em 2018 IEEE Spoken Language Technology Workshop (SLT)}, 2018,
  pp. 558--565.

\bibitem{bu2017aishell}
Hui Bu, Jiayu Du, Xingyu Na, Bengu Wu, and Hao Zheng,
\newblock ``Aishell-1: An open-source mandarin speech corpus and a speech
  recognition baseline,''
\newblock in {\em 2017 20th conference of the oriental chapter of the
  international coordinating committee on speech databases and speech I/O
  systems and assessment (O-COCOSDA)}. IEEE, 2017, pp. 1--5.

\bibitem{dubey2023icassp}
Harishchandra Dubey, Ashkan Aazami, Vishak Gopal, Babak Naderi, Sebastian
  Braun, Ross Cutler, Alex Ju, Mehdi Zohourian, Min Tang, Hannes Gamper,
  et~al.,
\newblock ``Icassp 2023 deep speech enhancement challenge,''
\newblock {\em arXiv preprint arXiv:2303.11510}, 2023.

\bibitem{scheibler2018pyroomacoustics}
Robin Scheibler, Eric Bezzam, and Ivan Dokmani{\'c},
\newblock ``Pyroomacoustics: A python package for audio room simulation and
  array processing algorithms,''
\newblock in {\em 2018 IEEE international conference on acoustics, speech and
  signal processing (ICASSP)}. IEEE, 2018, pp. 351--355.

\bibitem{gu2020multi}
Rongzhi Gu, Shi-Xiong Zhang, Yong Xu, Lianwu Chen, Yuexian Zou, and Dong Yu,
\newblock ``Multi-modal multi-channel target speech separation,''
\newblock {\em IEEE Journal of Selected Topics in Signal Processing}, vol. 14,
  no. 3, pp. 530--541, 2020.

\bibitem{bai2018empirical}
Shaojie Bai, J~Zico Kolter, and Vladlen Koltun,
\newblock ``An empirical evaluation of generic convolutional and recurrent
  networks for sequence modeling,''
\newblock {\em arXiv preprint arXiv:1803.01271}, 2018.

\bibitem{schroter2022deepfilternet}
Hendrik Schroter, Alberto~N Escalante-B, Tobias Rosenkranz, and Andreas Maier,
\newblock ``Deepfilternet: A low complexity speech enhancement framework for
  full-band audio based on deep filtering,''
\newblock in {\em ICASSP 2022-2022 IEEE International Conference on Acoustics,
  Speech and Signal Processing (ICASSP)}. IEEE, 2022, pp. 7407--7411.

\bibitem{8683855}
Jonathan~Le Roux, Scott Wisdom, Hakan Erdogan, and John~R. Hershey,
\newblock ``Sdr – half-baked or well done?,''
\newblock in {\em ICASSP 2019 - 2019 IEEE International Conference on
  Acoustics, Speech and Signal Processing (ICASSP)}, 2019, pp. 626--630.

\bibitem{941023}
A.W. Rix, J.G. Beerends, M.P. Hollier, and A.P. Hekstra,
\newblock ``Perceptual evaluation of speech quality (pesq)-a new method for
  speech quality assessment of telephone networks and codecs,''
\newblock in {\em 2001 IEEE International Conference on Acoustics, Speech, and
  Signal Processing. Proceedings (Cat. No.01CH37221)}, 2001, vol.~2, pp.
  749--752 vol.2.

\bibitem{5495701}
Cees~H. Taal, Richard~C. Hendriks, Richard Heusdens, and Jesper Jensen,
\newblock ``A short-time objective intelligibility measure for time-frequency
  weighted noisy speech,''
\newblock in {\em 2010 IEEE International Conference on Acoustics, Speech and
  Signal Processing}, 2010, pp. 4214--4217.

\bibitem{gao2022paraformer}
Zhifu Gao, Shiliang Zhang, Ian McLoughlin, and Zhijie Yan,
\newblock ``Paraformer: Fast and accurate parallel transformer for
  non-autoregressive end-to-end speech recognition,''
\newblock in {\em INTERSPEECH}, 2022.

\end{thebibliography}

\end{document}